\documentstyle[12pt,epsf]{article}
\def\fo{\hbox{{1}\kern-.25em\hbox{l}}}

\newcommand{\newc}{\newcommand}

\newc{\lcal}{\int {\cal L}dt}

\newc{\LSP}{{\chi^0_1}}
\newc{\stauR}{{\tilde \tau_R}}
\newc{\stau}{{\tilde \tau_1}}
\newc{\mstop}{m_{\tilde{t}}}
\newc{\mHpm}{m_{H^\pm}}
\newc{\gsim}{\lower.7ex\hbox{$\;\stackrel{\textstyle>}{\sim}\;$}}
\newc{\lsim}{\lower.7ex\hbox{$\;\stackrel{\textstyle<}{\sim}\;$}}
\newc{\ie}{{\it i.e.}}
\newc{\etal}{{\it et al.}}
\newc{\eg}{{\it e.g.}}
\newc{\kev}{\hbox{\rm\,keV}}
\newc{\mev}{\hbox{\rm\,MeV}}
\newc{\gev}{\hbox{\rm\,GeV}}
\newc{\tev}{\hbox{\rm\,TeV}}
\newc{\xpb}{\hbox{\rm\, pb}}
\newc{\xfb}{\hbox{\rm\, fb}}

\newc{\mtop}{m_t}
\newc{\mbot}{m_b}
\newc{\mz}{m_Z}
\newc{\mw}{M_W}
\newc{\alphasmz}{\alpha_s(m_Z^2)}
\newc{\swsq}{\sin^2\theta_W}
\newc{\tw}{\tan\theta_W}
\newc{\cw}{\cos\theta_W}
\newc{\sw}{\sin\theta_W}
\newc{\BR}{\hbox{\rm BR}}
\newc{\zbb}{Z\to b\bar}
\newc{\Gb}{\Gamma (Z\to b\bar b)}
\newc{\Gh}{\Gamma (Z\to \hbox{\rm hadrons})}
\newc{\rbsm}{R_b^\hbox{\rm sm}}
\newc{\rbsusy}{R_b^\hbox{\rm susy}}
\newc{\drb}{\delta R_b}

\newc{\sgn}{\mbox{sgn}}

\newc{\tbeta}{\tan\beta}
\newc{\uL}{{\tilde u_L}}
\newc{\uR}{{\tilde u_R}}
\newc{\cL}{{\tilde c_L}}
\newc{\cR}{{\tilde c_R}}
\newc{\tL}{{\tilde t_L}}
\newc{\tR}{{\tilde t_R}}
\newc{\dL}{{\tilde d_L}}
\newc{\dR}{{\tilde d_R}}
\newc{\sL}{{\tilde s_L}}
\newc{\sR}{{\tilde s_R}}
\newc{\bL}{{\tilde b_L}}
\newc{\bR}{{\tilde b_R}}
\newc{\eL}{{\tilde e_L}}
\newc{\eR}{{\tilde e_R}}
\newc{\mhp}{m_{H^\pm}}
\newc{\mhalf}{m_{1/2}}
\newc{\emt}{{e/\mu /\tau}}

\newc{\lR}{\tilde{l}_R}
\newc{\lL}{\tilde{l}_L}
\newc{\nL}{\tilde{\nu}_L}
\newc{\na}{\chi^0_1}
\newc{\nb}{\chi^0_2}
\newc{\nc}{\chi^0_3}
\newc{\nd}{\chi^0_4}
\newc{\ca}{\chi^{\pm}_1}
\newc{\cb}{\chi^{\pm}_2}
\newc{\camp}{\chi^\mp_1}
\newc{\cbmp}{\chi^\mp_1}
\newc{\capos}{\chi^{+}_1}
\newc{\caneg}{\chi^{-}_1}
\newc{\phit}{\phi_t}
\newc{\phib}{\phi_b}
\newc{\phiew}{\phi_{ew}}
\newc{\htz}{h^0_t}
\newc{\hbz}{h^0_b}
\newc{\hewz}{h^0_{ew}}
\newc{\hsmz}{h^0_{sm}}
\newc{\huz}{h^0_u}
\newc{\hsusyz}{h^0_{susy}}

\def\mp{M_P}

\def\beq{\begin{equation}}
\def\eeq{\end{equation}}
\def\bea{\begin{eqnarray}}
\def\eea{\end{eqnarray}}
\def\slashchar#1{\setbox0=\hbox{$#1$}           
   \dimen0=\wd0                                 
   \setbox1=\hbox{/} \dimen1=\wd1               
   \ifdim\dimen0>\dimen1                        
      \rlap{\hbox to \dimen0{\hfil/\hfil}}      
      #1                                        
   \else                                        
      \rlap{\hbox to \dimen1{\hfil$#1$\hfil}}   
      /                                         
   \fi}                                         %
\catcode`@=11
\long\def\@caption#1[#2]#3{\par\addcontentsline{\csname
  ext@#1\endcsname}{#1}{\protect\numberline{\csname
  the#1\endcsname}{\ignorespaces #2}}\begingroup
    \small
    \@parboxrestore
    \@makecaption{\csname fnum@#1\endcsname}{\ignorespaces #3}\par
  \endgroup}
\catcode`@=12

\begin{document}

\baselineskip=18pt

\begin{titlepage}

\begin{center}
\vspace{1cm}

{\Large \bf Effective Hamiltonian for non-minimally coupled scalar
fields}

\vspace{0.5cm}

{\bf Emine Me{\c s}e$^{1}$, 
Nurettin Pirin{\c c}{\c c}io{\~g}lu$^{1}$, 
Irfan A{\c c}{\i}kg{\"o}z$^{1}$ and 
Figen Binbay$^{1}$}  

\vspace{.8cm} {\it $^1$ Department of Physics, Dicle University,
TR21280, Diyarbak{\i}r, Turkey}

\end{center}
\vspace{1cm}

\begin{abstract}
\medskip
Performing a relativistic approximation as the generalization to a curved 
spacetime of the flat space Klein-Gordon equation, an effective Hamiltonian 
which includes non-minimial coupling between gravity and scalar field and also 
quartic self-interaction of scalar field term is obtained.  


\end{abstract}

\end{titlepage}


\section{Introduction}

The Hamilton operator for quantum optics in gravitational fields is derived 
using minimal coupling and obtained that electric field as well as the dipole 
are operationally defined by measured quantities.  In contrast, the magnetic 
dipole coupling can be modified by 
the gravitational field 
\cite{lammerzahl95}.  
This effect 
is important for the interaction of mesoscopic quantum system with 
gravitational fields.  
An intensive review on interaction of mesoscopic systems 
including an overview of classical gravitaitonal waves is given by 
Kiefer and Weber 
\cite{Kiefer2004}.  
There are many other works can be given which explored these effects 
\cite{Barcelo2003}, 
\cite{Marshall2003}, 
\cite{Hehl2003}.  
The gravitational interaction is distiguished by the 
fact that it interacts universally with all forms of energy.  It dominates 
on a large scales (cosmology) and for compact objects (such as neutron 
stars and black holes).  In large scales in curved spaces nonminimal 
coupling term is to be expected
\cite{Faroni98}, 
\cite{Gunzig2000}.  

In this work generalization to a curved spacetime of the flat space 
Klein-Gordon (KG) equation is considered and quartic self-interaction term in addition to nonminimal coupling effects is included.  The motivation for 
chosing a quartic self-interaction comes from the followings:  
Energy of an homogenous superconductor is charecterised by Ginzburg-Landau 
model 
\cite{Kiefer2004}
\cite{bcs}
\cite{landau}
\cite{gorkov}
\cite{Muller97}.  
This is exactly the same as the quartic self-
interaction term.  

The following section represent the derived Hamiltonian for the KG equation 
in curved space time including quartic self-interaction term, and the 
paper is completed with a short conclusion.  

\section{Non-minimally coupled scalar fields}
We consider a generic complex scalar field $\phi(x)$ with
squared-mass $m^2$ and quartic coupling $\lambda$. Its equation of
motion is given by
\begin{eqnarray}
\label{eom} g^{\mu \nu} \nabla_{\mu} \nabla_{\nu} \phi - \frac{m^2
c^2}{\hbar^2} \phi - \zeta {\cal{R}} \phi - \frac{\lambda}{\hbar^2
c^2} \left|\phi\right|^2 \phi = 0
\end{eqnarray}
where the dimensionless parameter $\zeta$ stands for the direct
coupling of $\phi$ to curvature scalar ${\cal{R}}$. The
derivatives in the first term are meant to be covariant with
respect to both general coordinate and gauge transformations:
\begin{eqnarray}
\nabla_{\mu} V^{\nu} \equiv \partial_{\mu} V^{\nu} +
\Gamma^{\nu}_{\mu \alpha} V^{\alpha} - i \frac{q_{\phi}}{\hbar c}
A_{\mu} V^{\nu}
\end{eqnarray}
for $V_{\mu} = \nabla_{\mu} \phi\equiv \partial_{\mu} \phi$ with
$q_{\phi}$ being the electric charge of $\phi$ and
$\Gamma^{\mu}_{\nu \lambda}$ the connection coefficients.

For a through understanding of the energetics and dynamics of
mesoscopic quantum systems represented by $\phi$, it suffices to
treat gravitational and electromagnetic fields as classical
backgrounds. In post-Newtonian approximation, the most general
parameterization of the gravitational field $g_{\mu \nu}(x)$
(created by entire matter and energy surrounding the mesoscopic
structure) is provided by the PPN formalism \cite{PPN}:
\begin{eqnarray}
\label{metric} g_{00} &=& - \left[ 1 - 2 \frac{U}{c^2} + 2 \beta
\left(
\frac{U}{c^2}\right)^2\right]\nonumber\\
g_{i j}&=& \left[ 1+ 2 \gamma \frac{U}{c^2}\right] \delta_{i j}
\end{eqnarray}
and $g_{0 i} =0$, for $\forall$ $i,j=1,2,3$. This parameterization
rests on the assumption that entire surrounding matter is at rest
as otherwise there would be a nonvanishing $g_{0 i}$ induced by
flow of matter. The parameters $\beta$ and $\gamma$ (each of which
equals unity in Einstein gravity) measure, respectively,
nonlinearity in gravitational interactions and amount of spacetime
curving induced by unit mass. Here $U$ is nothing but the
instantaneous Newton potential
\begin{eqnarray}
U(\vec{x}, t)= G_N \int d^{3} \vec{x}^{\prime}\,
\frac{\rho_{m}(\vec{x}^{\prime}, t)}{\left|\vec{x} -
\vec{x}^{\prime} \right|}
\end{eqnarray}
where $\rho_{m}(\vec{x}^{\prime}, t)$ is the rest mass density of
the matter distribution generating $g_{\mu \nu}(x)$ above.

After inserting (\ref{metric}) in (\ref{eom}), and going to
non-relativistic limit (by discarding high-frequency modes with
$\omega = m c^2/\hbar$ as in \cite{lammerzahl95}) the equation of
motion (\ref{eom}) can be recast into the form
\begin{eqnarray}
\label{sch} i \hbar \partial_{t} \varphi = H \varphi
\end{eqnarray}
where $\varphi(x) = e^{ \frac{i}{\hbar} m c^2 t } \phi(x)$, and
\begin{eqnarray}
\label{H_we}
H  &=& \left(1 - \frac{ \vec{p}^2}{4 m^2 c^2} - (2 \gamma +1)
\frac{U}{c^2} \right) \frac{\vec{p}^2}{2 m}  - q_{\phi} A_{0} - m
U - \left( \frac{1}{2} - \beta \right) m \frac{U^2}{c^2} - i \hbar
\frac{(2 \gamma +1)}{2 m c^2} \vec{g} \cdot \vec{p}\nonumber\\ &-&
\frac{3 \pi G_{N} \hbar^2}{m c^2} \left(\gamma - \frac{4}{3}
\left( 2 \gamma -1\right) \zeta\right) \rho_{m}  +
\frac{\lambda}{2 m c^2} \left| \varphi \right|^2
\end{eqnarray}
with $\vec{g} = - \vec{\nabla} U$ being the gravitational
acceleration, and $\vec{p} = - i \hbar \left( \vec{\nabla} -
\frac{i q_{\phi}}{\hbar c} \vec{A}\right)$ the mechanical
momentum.

The Hamiltonian $H$ given in 
(\ref{H_we}) 
is hermitian with respect to the Sch\"odinger scalar 
product.  All time-dependent hermiticity-violating terms are canceled due to 
the time dependent transformation as it is done in 
\cite{lammerzahl95}.  
This Hamiltonian includes the kinetic energy terms of the system to the first 
relativistic correction to the coupling of the matter field to the gravity.  
The term represents the coupling of the matter includes gravitational Darwin 
term.  It is clear to see that the gravitational Darwin term is modified by 
the influence of nonminimal coupling with comparing the Hamiltonian $H$ 
with those obtained by Lammerzahl in 
\cite{lammerzahl95}.  
In addition Equation 
(\ref{H_we}) 
includes the quartic self-interaction term which can be considered as energy term for superconductor charecterised by Landau-Ginzburg model.  

\section{Conclusion}

In this study we have shown that nonminimal coupling effects modifies the 
gravitational Darwin term.  In addition quartic self-interaction 
is represented in the Hamiltonian.  
Since quartic self-interaction term in Hamiltonian 
(\ref{H_we}) 
is similar to the energy expansion of an homogenous superconductor 
charecterised by Ginzburg-Landau model, a future work of exploring 
gravitaitonal field effects on the critical temprature $T_c$ of 
superconductor can be suggested.  

\section*{Acknowledgements}
The authors would like to thank to D. A. Demir for the suggestion of 
the research proposal and useful discussion.


\begin{thebibliography}{99}

\bibitem{lammerzahl95}
C.~Laemmerzahl, 
{\it Phys. Lett. A} {\bf 203}, 12 (1995).


\bibitem{Kiefer2004}
C. Kiefer and C. Weber, 
{\it arXiv:gr-qc/0408010} {\bf v2}, (2004).

\bibitem{Barcelo2003}
C. Barcelo, S. Liberati and M. Visser, 
{\it Phys. Rev. A} {\bf 68}, 053613 (2003).

\bibitem{Marshall2003}
W. Marshall, C. Simon, R. Penrose and D. Bouwmeester,
{\it Phys. Rev Lett.} {\bf 91}, 130401 (2003).

\bibitem{Hehl2003}
F. W. Hehl, Y. N. Obukhov and B. Rosenow, 
{\it Phys. Rev Lett.} {\bf 93}, 096804 (2004).

\bibitem{Faroni98}
V. Faraoni, E. Gunzig and P. Nardone, 
{\it arXiv:gr-qc/9811047} {\bf v1}, (1998).

\bibitem{Gunzig2000}
E. Gunzig et al.,  
{\it Int. J. of Theor. Phys.} {\bf 39}, 1901 (2000).

\bibitem{bcs}
J.~Bardeen, L.~N.~Cooper and J.~R.~Schrieffer, 
{\it Phys. Rev.} {\bf 108} (1957) 1175.

\bibitem{landau}
V.~L.~Ginzburg and L.~D.~Landau, 
{\it Zh. Eksp. Teor. Fiz.}  {\bf 20} (1950) 1064.

\bibitem{gorkov}
L.~P.~Gor'kov, {\it Sov. Phys. JETP} {\bf 8} (1958) 505; 
{\it Sov.
Phys. JETP}  {\bf 9} (1959) 1364.

\bibitem{Muller97}
P. Muller and A. V. Ustnov, 
{\it The Physics of Superconductivity}, 
Springer-Verlag, Berlin Heidelberg (1997).  

\bibitem{PPN}
C.~M.~Will, 
{\it Living Rev. Rel.}  {\bf 4} (2001) 4 
{\it arXiv:gr-qc/0103036}.  


\end{thebibliography}
\end{document}